\documentstyle[12pt]{article}
\begin{document}

\def\1op{\hbox{{\rm 1}\kern -0.23em\hbox{{\rm I}}}}

\title{ON AXIALLY SYMMETRIC SOLUTIONS\\ IN THE ELECTROWEAK THEORY}
\vspace{1.5truecm}
\author{
{\bf Yves Brihaye}\\
Facult\'e des Sciences, Universit\'e de Mons-Hainaut\\
B-7000 Mons, Belgium
\and
{\bf Jutta Kunz}\\
Fachbereich Physik, Universit\"at Oldenburg, Postfach 2503\\
D-26111 Oldenburg, Germany\\
and\\
Instituut voor Theoretische Fysica, Rijksuniversiteit te Utrecht\\
NL-3508 TA Utrecht, The Netherlands}

\vspace{1.5truecm}

\date{March 30, 1994}

\maketitle
\vspace{1.0truecm}

\begin{abstract}

We present the general ansatz,
the energy density and the Chern-Simons charge
for static axially symmetric configurations
in the bosonic sector of the electroweak theory.
Containing the sphaleron, the multisphalerons
and the sphaleron-antisphaleron pair at finite mixing angle,
the ansatz further allows the construction
of the sphaleron and multisphaleron barriers
and of the bisphalerons at finite mixing angle.
We conjecture that further solutions exist.

\end{abstract}
\vfill   \noindent Univ. Mons-Hainaut-Preprint \hfill \break
         \noindent Univ. Utrecht-Preprint THU-94/7
\vfill\eject

\section{Introduction}

In the electroweak theory several types
of classical solutions are known.
A decade ago the sphaleron solution of the electroweak theory
was discovered \cite{Man1}
and constructed in the limit of vanishing mixing angle
\cite{KM}. In this limit the sphaleron is
spherically symmetric and parity reflection symmetric.
Much later the sphaleron was constructed for the
full electroweak theory with gauge group SU(2)$\otimes$U(1)
\cite{KKB1,KKB2}.
At finite mixing angle the sphaleron is only axially symmetric,
but it retains its parity reflection symmetry.
At the physical mixing angle
the spherical approximation for the sphaleron is excellent
\cite{KKB1,KKB2}.

Recently further solutions of the
electroweak theory have been constructed,
which are axially symmetric and symmetric under parity reflections.
These are on the one hand the multisphaleron solutions
\cite{KK} and on the other hand the sphaleron--antisphaleron pair
\cite{Kl1,Kl2}.
The multisphaleron solutions carry
Chern-Simons charge $N_{\rm CS}=n/2$,
where $n$ is an integer counting the winding of the fields
in the azimuthal angle $\phi$.
The sphaleron has winding number $n=1$.
Like the sphaleron the multisphalerons
are thus associated with fermion number violation \cite{KK}.
In contrast the sphaleron--antisphaleron pair carries
Chern-Simons charge $N_{\rm CS}=0$ \cite{Kl2}.
The ansatz for the sphaleron--antisphaleron pair
can be generalized by realizing,
that it involves a winding of the fields in the angle $\theta$.
Denoting the corresponding winding number $m$,
the sphaleron--antisphaleron pair has
winding number $m=2$, while the sphaleron has $m=1$.

When constructing non-contractible loops
in configuration space, the intermediate configurations
between the vacua and the sphaleron,
representing the sphaleron barrier,
have less symmetries than the sphaleron,
even for vanishing mixing angle \cite{Man1,AKY}.
Indeed in the limit of vanishing mixing angle
the construction of the sphaleron barrier
involves configurations which do not retain
the discrete symmetry of the sphaleron, parity reflection symmetry.
For finite mixing angle the sphaleron barrier
has not yet been constructed.

Furthermore, for high values of the Higgs mass
new classical solutions appear in the electroweak theory,
the bisphalerons \cite{KB,LGY}.
These solutions, constructed so far only for vanishing mixing angle,
where they are spherically symmetric,
are not invariant under parity, but occur as parity doublets.
Like the sphaleron,
at finite mixing angle they will retain only axial symmetry.
The bisphalerons are lower in energy than the sphaleron
\cite{KB,LGY}.
This was demonstrated also in a perturbative analysis for finite
mixing angle \cite{BK}.
Therefore at large Higgs masses the lowest bisphalerons
represent the top of the energy barrier between
neighbouring topologically distinct vacua.
The construction of the bisphalerons at finite
mixing angle is an outstanding problem.

In this paper we develop the formalism for the construction
of general classical static configurations of the electroweak theory
with axial symmetry.
In section 2 we present the general ansatz for the fields
and the energy density obtained with this ansatz.
Further we discuss the four residual gauge symmetries
of the energy density and several choices of gauge.
In section 3 we discuss the classical solutions
obtainable with this ansatz,
the sphaleron and multisphalerons,
the sphaleron-antisphaleron pair and their generalizations
as well as the bisphalerons and the barriers.
In section 4 we present the Chern-Simons charge
for the general ansatz.
Further we evaluate the Chern-Simons charge
for the multisphalerons and for the solutions
which may be obtained with the
generalized sphaleron-antisphaleron pair ansatz.
We present our conclusions in section 5.

\section{\bf Ansatz and energy density}

Let us consider the bosonic sector of the Weinberg-Salam theory
\begin{equation}
{\cal L} = -\frac{1}{4} F_{\mu\nu}^a F^{\mu\nu,a}
- {1 \over 4} f_{\mu \nu} f^{\mu \nu}
+ (D_\mu \Phi)^{\dagger} (D^\mu \Phi)
- \lambda (\Phi^{\dagger} \Phi - \frac{v^2}{2} )^2
\   \end{equation}
with the SU(2) field strength tensor
\begin{equation}
F_{\mu\nu}^a=\partial_\mu W_\nu^a-\partial_\nu W_\mu^a
            + g \epsilon^{abc} W_\mu^b W_\nu^c
\ , \end{equation}
with the U(1) field strength tensor
\begin{equation}
f_{\mu\nu}=\partial_\mu A_\nu-\partial_\nu A_\mu
\ , \end{equation}
and the covariant derivative for the Higgs field
\begin{equation}
D_{\mu} \Phi = \Bigl(\partial_{\mu}
             -\frac{i}{2}g \tau^a W_{\mu}^a
             -\frac{i}{2}g' A_{\mu} \Bigr)\Phi
\ . \end{equation}
The gauge symmetry is spontaneously broken
due to the non-vanishing vacuum expectation
value $v$ of the Higgs field
\begin{equation}
    \langle \Phi \rangle = \frac{v}{\sqrt2}
    \left( \begin{array}{c} 0\\1  \end{array} \right)
\ , \end{equation}
leading to the boson masses
\begin{equation}
    M_W = \frac{1}{2} g v \ , \ \ \ \
    M_Z = {1\over2} \sqrt{(g^2+g'^2)} v \ , \ \ \ \
    M_H = v \sqrt{2 \lambda}
\ . \end{equation}
The mixing angle $\theta_{\rm w}$ is determined by
the relation $ \tan \theta_{\rm w} = g'/g $,
and the electric charge is $e = g \sin \theta_{\rm w}$.

\subsection{Axially symmetric ansatz}

Let us introduce the set of orthonormal vectors \cite{Man2,RR,KK}
\begin{eqnarray}
\vec u_1^{(n)}(\phi) & = & (\cos n \phi, \sin n \phi, 0) \ ,
\nonumber \\
\vec u_2^{(n)}(\phi) & = & (0, 0, 1) \ ,
\nonumber \\
\vec u_3^{(n)}(\phi) & = & (\sin n \phi, - \cos n \phi, 0)
\ , \end{eqnarray}
and the matrices
\begin{equation}
G_i^{(n)}(\phi) = u_i^{a(n)}(\phi) \tau^a
\ , \end{equation}
where $\tau^a$ are the Pauli matrices and
$\phi$  is the azimuthal angle defined via
\begin{equation}
(x, y, z) = (\rho \cos \phi, \rho \sin \phi, z) =
 (r \sin \theta \cos \phi, r \sin \theta \sin \phi, r \cos \theta)
\ . \end{equation}
The static axially symmetric ansatz for
the SU(2) gauge fields, the U(1) gauge field
and the Higgs field is then given by
\begin{eqnarray}
     W_i(\vec r\,) &=& W_i^a(\vec r\,) \tau^a
                    = u_j^{i(1)}(\phi) G_k^{(n)}(\phi) w_j^k(\rho, z) \
, \nonumber\\
     W_0(\vec r\,) &=& W_0^a(\vec r\,) \tau^a = 0
\ , \end{eqnarray}
\begin{eqnarray}
     A_i(\vec r\,) &=& u_j^{i(1)}(\phi) a_j(\rho , z)
, \nonumber\\
     A_0(\vec r\,) &=& 0
\ , \end{eqnarray}
\begin{equation}
    \Phi(\vec r\,) = \frac{v}{\sqrt2}
   \Bigl( h_0(\rho,z) + ih_j(\rho,z) G_j^{(n)} \Bigr)
    \left( \begin{array}{c} 0\\1  \end{array} \right)
\ , \end{equation}
where the indices $i,j,k$ and $a$ run from 1 to 3.

This ansatz contains 16 arbitrary real functions of the variables
$\rho$ and $z$.
The ansatz is axially symmetric,
i.~e.~a rotation around the z-axis can be compensated
by a gauge transformation.
(For the Higgs field the compensating gauge transformation is an
element of the diagonal group U(1)$\oplus$U(1),
the first U(1) being the subgroup
of SU(2) generated by the matrix $G_2^{(n)}$.)

\subsection{Energy functional}

The resulting axially symmetric energy functional $E$
\begin{equation}
E = \frac{1}{2} \int ( E_w + E_a + v^2 E_h )
	      \, d\phi \, \rho d\rho \, dz
\   \end{equation}
then has the contributions
\begin{eqnarray}
E_w &=& (\partial_{\rho} w^1_3 + {1\over {\rho}} (nw^3_1+w^1_3)
                - g(w^2_3 w^3_1-w^3_3w^2_1))^2 \nonumber \\
    &+& (\partial_{\rho} w^2_3 + {1\over {\rho}} w^2_3
                - g(w^1_1 w^3_3 - w^1_3w^3_1))^2\nonumber \\
    &+&(\partial_{\rho} w^3_3+{1\over {\rho}} (w^3_3-nw^1_1)
                - g(w^1_3w^2_1-w^1_1w^2_3))^2 \nonumber \\
    &+&(\partial_z w^1_3 + n{w^3_2\over {\rho}}
                - g(w^2_3w^3_2-w^3_3w^2_2))^2\\
    &+&(\partial_z w^2_3 - g(w_3^3w^1_2-w^3_2w^1_3))^2\nonumber \\
    &+&(\partial_z w^3_3- n {w_2^1\over {\rho}}
                 - g(w^2_2 w^1_3-w^2_3w^1_2))^2\nonumber \\
    &+&(\partial_{\rho}w^3_2 - \partial_z w^3_1 -
               g(w^2_1w^1_2-w^1_1w^2_2))^2 \nonumber \\
    &+&(\partial_z w^2_1 - \partial_{\rho} w^2_2  -
               g(w^3_1w^1_2-w^1_1w^3_2))^2\nonumber \\
    &+&(\partial_z w^1_1 - \partial_{\rho} w^1_2
              - g(w^3_2w^2_1-w^2_2w^3_1))^2 \nonumber
\ , \end{eqnarray}
\begin{equation}
E_a = (\partial_\rho a_3 + {1\over{\rho}} a_3 )^2
       + (\partial_z a_3 )^2
       + (\partial_\rho a_2 - \partial_z a_1)^2
\ , \end{equation}
\begin{eqnarray}
E_h  &=&   (  \partial_{\rho} h_0
             + {g  \over 2} (h_1 w_1^1 + h_2 w_1^2 + h_3 w_1^3)
             - {g' \over 2} (h_2 a_1)  )^2   \nonumber  \\
     &+&  (  \partial_{\rho} h_1
             + {g  \over 2} (h_3 w_1^2 - h_2 w_1^3 - h_0 w_1^1)
             + {g' \over 2} (h_3 a_1)  )^2   \nonumber  \\
     &+&  (  \partial_{\rho} h_2
             + {g  \over 2} (h_1 w_1^3 - h_3 w_1^1 - h_0 w_1^2)
             + {g' \over 2} (h_0 a_1)  )^2   \nonumber \\
    &+&  (  \partial_{\rho} h_3
             + {g  \over 2} (h_2 w_1^1 - h_1 w_1^2 - h_0 w_1^3)
             - {g' \over 2} (h_1 a_1)  )^2     \nonumber \\
    &+&  (  \partial_z h_0
             + {g  \over 2} (h_1 w_2^1 + h_2 w_2^2 + h_3 w_2^3)
             - {g' \over 2} (h_2 a_2)  )^2     \nonumber\\
     &+&  (  \partial_z h_1
             + {g  \over 2} (h_3 w_2^2 - h_2 w_2^3 - h_0 w_2^1)
             + {g' \over 2} (h_3 a_2)  )^2 \\
     &+&  (  \partial_z h_2
             + {g  \over 2} (h_1 w_2^3 - h_3 w_2^1 - h_0 w_2^2)
             + {g' \over 2} (h_0 a_2)  )^2  \nonumber  \\
    &+&  (  \partial_z h_3
             + {g  \over 2} (h_2 w_2^1 - h_1 w_2^2 - h_0 w_2^3)
             - {g' \over 2} (h_1 a_2)  )^2   \nonumber \\
    &+& (   {g \over 2} (h_1 w_3^3 - h_0 w_3^2 - h_3 w_3^1)
          + {g'\over 2} h_0 a_3   )^2        \nonumber   \\
    &+& (   {g \over 2} (h_0 w_3^3 - h_2 w_3^1 + h_1 w_3^2)
          + {g'\over 2} h_1 a_3  - {n h_1 \over {\rho}}  )^2  \nonumber \\
    &+& (   {g \over 2} (-h_1 w_3^1 - h_2 w_3^2 - h_3 w_3^3)
          + {g'\over 2} h_2 a_3   )^2  \nonumber \\
   &+& (   {g \over 2} (h_3 w_3^2 - h_0 w_3^1 - h_2 w_3^3)
          + {g'\over 2} h_3 a_3    - {n h_3 \over {\rho}}  )^2 \nonumber \\
   &+& { {\lambda v^2}\over2} ( h_0^2 + h_1^2 + h_2^2  + h_3^2 - 1 )^2
   \nonumber
\ . \end{eqnarray}

\subsection{Residual gauge symmetries}

The energy functional is invariant under a large class of gauge
transformations, which keep the ansatz forminvariant.
These gauge transformations are given by
\begin{eqnarray}
 U_0(\vec r\,) &=& \exp \Bigl(i\Gamma_0(\rho,z)\Bigr) \ , \nonumber \\
 U_1(\vec r\,) &=& \exp \Bigl(i\Gamma_1(\rho,z) G_1^{(n)}(\phi)\Bigr) \ ,
\nonumber \\
 U_2(\vec r\,) &=& \exp \Bigl(i\Gamma_2(\rho,z) G_2^{(n)}(\phi)\Bigr) \ ,
\nonumber \\
 U_3(\vec r\,) &=& \exp \Bigl(i\Gamma_3(\rho,z) G_3^{(n)}(\phi)\Bigr)
\ . \end{eqnarray}

\subsubsection{Transformation properties of the fields}

Considering first the transformation $U_0$,
the components of the abelian gauge field $a_i$ transform as
\begin{equation}
  a_1' = a_1 + {2 \over g'}
 {\partial \Gamma_0 \over \partial \rho} \ , \quad
   a_2' = a_2 + {2 \over g'} {\partial \Gamma_0 \over \partial z}
      \ , \quad  a_3' = a_3
\ , \end{equation}
the Higgs field components
\begin{equation}
    \left( \begin{array}{c} h_3\\h_1  \end{array} \right) \ , \ \ \
    \left( \begin{array}{c} h_0\\h_2  \end{array} \right)
\   \end{equation}
transform as doublets with angle $\Gamma_0$,
and the SU(2) fields are invariant under $U_0$.

Considering next the three abelian transformations $U_i$ generated by
$G_i^{(n)}$,
the abelian gauge field is invariant under these transformations,
while the components of the non-abelian gauge fields
$w_a^b$ and of the Higgs field $h_0$ and $h_i$ form various multiplets.

Under the transformation $U_3 = \exp (i \Gamma_3 G_3^{(n)})$
the SU(2) gauge field components
\begin{equation}
    \left( \begin{array}{c} w_1^1\\w_1^2  \end{array} \right) \ , \ \ \
    \left( \begin{array}{c} w_2^1\\w_2^2  \end{array} \right) \ , \ \ \
    \left( \begin{array}{c} w_3^1\\w_3^2 -\frac{n}{g \rho}
         \end{array} \right)
\   \end{equation}
transform as doublets with angle $2 \Gamma_3$,
$w_3^3$ is invariant and the two remaining components
$(w_1^3 , w_2^3)$ transform as a two dimensional
gauge field
\begin{equation}
  \Bigl(w_1^3\Bigr)'   = w_1^3
      + {2 \over g} {\partial \Gamma_3 \over \partial \rho} \ , \quad
  \Bigl(w_2^3\Bigr)'   = w_2^3
      + {2 \over g} {\partial \Gamma_3 \over \partial z}
\ . \end{equation}
The Higgs field components
\begin{equation}
    \left( \begin{array}{c} h_1\\h_2  \end{array} \right) \ , \ \ \
    \left( \begin{array}{c} h_3\\h_0  \end{array} \right)
\   \end{equation}
transform as doublets with angle $\Gamma_3$.

Analogously,
under the transformations $U_1 = \exp(i \Gamma_1 G_1^{(n)})$
and $U_2 = \exp (i \Gamma_2 G_2^{(n)})$ similar schemes occur with
\begin{equation}
    \left( \begin{array}{c} w_1^2\\w_1^3  \end{array} \right) \ , \ \ \
    \left( \begin{array}{c} w_2^2\\w_2^3  \end{array} \right) \ , \ \ \
    \left( \begin{array}{c} w_3^2 -\frac{n}{g \rho}\\w_3^3
         \end{array} \right) \ , \ \ \
    w_3^1 \ , \ \ \
    \left( \begin{array}{c} w_1^1\\w_2^1  \end{array} \right)
\ , \end{equation}
and
\begin{equation}
    \left( \begin{array}{c} w_1^3\\w_1^1  \end{array} \right) \ , \ \ \
    \left( \begin{array}{c} w_2^3\\w_2^1  \end{array} \right) \ , \ \ \
    \left( \begin{array}{c} w_3^3\\w_3^1  \end{array} \right) \ , \ \ \
    w_3^2 \ , \ \ \
    \left( \begin{array}{c} w_1^2\\w_2^2  \end{array} \right)
\ , \end{equation}
respectively, for the SU(2) gauge field components,
and
\begin{equation}
    \left( \begin{array}{c} h_1\\h_0  \end{array} \right) \ , \ \ \
    \left( \begin{array}{c} h_2\\h_3  \end{array} \right)
\ , \end{equation}
and
\begin{equation}
    \left( \begin{array}{c} h_2\\h_0  \end{array} \right) \ , \ \ \
    \left( \begin{array}{c} h_3\\h_1  \end{array} \right)
\ , \end{equation}
respectively, for the Higgs field components.

\subsubsection{Choices of gauge}

In order to construct classical solutions, the four residual gauge degrees
of freedom need to be fixed.
There appear to be many different ways to fix these four
gauge degrees of freedom.
However, from our experience in constructing the
sphaleron at finite mixing angle, we know that care
must be taken to choose a gauge,
where the classical solutions are regular \cite{KKB1,KKB2,BKK}.

{\it Coulomb gauges}

For the single residual gauge degree of freedom present
for the sphaleron at finite mixing angle, $U_3$,
we chose the Coulomb gauge for the 2-dimensional gauge field
\begin{equation}
 {\partial  w_1^3 \over \partial \rho} +
    {\partial  w_2^3 \over \partial z } = 0
\ , \end{equation}
since it lead to regular classical solutions
\cite{KKB1,KKB2,BKK}.

We therefore suggest as a probably good choice of gauge
the Coulomb gauge for all four 2-dimensional gauge fields,
i.~e.~in addition to Eq.~(27)
\begin{equation}
 {\partial  w_1^1 \over \partial \rho} +
    {\partial  w_2^1 \over \partial z } = 0  \ , \ \ \ \
 {\partial  w_1^2 \over \partial \rho} +
    {\partial  w_2^2 \over \partial z } = 0  \ , \ \ \ \
 {\partial  a_1 \over \partial \rho} +
    {\partial  a_2 \over \partial z } = 0
\ . \end{equation}
In the general case such a choice of gauge leaves
16 unknown functions to be determined numerically.

{\it Other gauges}

Another way of fixing the gauge consists of eliminating
one or more functions, leaving a smaller number of
unknown functions to be determined numerically.
Appearing attractive at first sight,
such gauge choices may prove to be singular \cite{BKK}.

Let us nevertheless consider such choices briefly.
For instance, setting the angular part
of the Higgs field in a canonical position,
we could obtain the {\it physical gauge}
or the {\it hedgehog gauge}.
In the {\it physical gauge} the Higgs field is specified
only by $h_0$, while $h_1=h_2=h_3=0$.
In the {\it hedgehog gauge} the Higgs field is specified
only by the function $h$, defined via
$h_1=h \sin \theta, \ h_2= h\cos \theta$,
while $h_0=h_3=0$.
Fixing three of the four degrees of freedom,
both these gauges are known to be singular for the sphaleron
at finite mixing angle \cite{BKK}.

Another possibly better choice could be to only assume
$h_3=0$ and supplement this gauge choice with the Coulomb gauge
for the remaining three degrees of freedom.
Note, that $h_3$ vanishes in all known classical solutions.

\section{Classical solutions}

All known static (3-dimensional) classical solutions
can be obtained from the general
static ansatz.
This ansatz further allows
to construct the sphaleron barrier at finite mixing angle,
to generalize the bisphalerons, known at vanishing mixing angle,
to finite mixing angle,
and to possibly construct new solutions.

\subsection{Barriers and bisphalerons}

Until now, vacuum to vacuum paths
passing the sphaleron have been constructed
only at vanishing mixing angle.
Since they involve parity violating configurations,
the general axially symmetric ansatz must be taken
to obtain such paths at finite mixing angle.

The general ansatz is also necessary
to obtain the barriers associated with multisphalerons,
as well as for the construction of bisphalerons at finite
mixing angle.

\subsubsection{Parametrization of the general ansatz}

In order to compare with the known spherical barrier,
and to take out the trivial angular dependence (on the angle $\theta$)
we parametrize the axial functions
in spherical coordinates as follows
\begin{eqnarray}
       w_1^3 &=& { 2 \over gr} F_1(r,\theta) \cos \theta \ , \ \ \ \ \
\ \ \ \ \ \ \ \
       w_2^3  =-{2 \over gr} F_2(r,\theta) \sin \theta \ ,
\nonumber \\
       w_3^1 &=&-{2n \over gr} F_3(r,\theta) \cos \theta \ , \ \ \ \ \
\ \ \ \ \
       w_3^2  = { 2n\over gr} F_4(r,\theta) \sin \theta \ ,
\nonumber \\
       w_1^2 &=& { 2 \over gr} H_1(r,\theta) \sin \theta \cos \theta  \ ,
\ \ \ \ \ \
       w_2^1  = { 2 \over gr} H_2(r,\theta) \sin \theta \cos \theta\ ,
\nonumber \\
       w_1^1 &=& { 2 \over gr}(H_3(r,\theta) \sin^2 \theta + H_4(r, \theta))
       \ ,
\nonumber \\
       w_2^2 &=& { 2 \over gr}(H_3(r,\theta) \cos^2 \theta + H_4(r, \theta))
       \ ,
\nonumber \\
       w_3^3 &=&  { 2n \over gr} H_5(r, \theta)\ ,
\nonumber \\
       h_1 &=& F_5(r, \theta) \sin \theta \ , \ \ \ \ \ \ \ \ \ \ \ \ \
\ \
       h_2 = F_6(r, \theta) \cos \theta \ ,
\nonumber \\
       h_3 &=& H_6(r, \theta) \sin \theta \ , \ \ \ \ \ \ \ \ \ \ \ \ \
\ \,
       h_0 = H_7(r, \theta) \ ,
\nonumber \\
  a_1 &=& {2 \over g'r} H_8(r,\theta) \sin \theta \cos \theta \ , \ \ \ \
  a_2  =  {2 \over g'r} H_9(r,\theta) \ ,
\nonumber \\
  a_3 &=& {2 \over g'r} F_7(r,\theta) \sin \theta
\ . \end{eqnarray}
This parametrization is a generalization of the parametrization
used for the sphaleron at finite mixing angle,
containing in addition to the seven functions $F_i(r,\theta)$
the nine functions $H_i(r,\theta)$.
The factors of $\sin \theta$ and $\cos \theta$
in the above parametrization
are chosen in accordance with the known spherical configurations,
the sphaleron, the sphaleron barrier and the bisphalerons,
where the functions $F_i(r,\theta)$ and $H_i(r,\theta)$
reduce to functions of the radial coordinate $r$ alone,
as shown below.

\subsubsection{Recovering spherical symmetry}

In the limit $\theta_{\rm w} =  0$
the sphaleron, the configurations along the sphaleron barrier
and the bisphalerons
are spherically symmetric.
The abelian gauge potential
can consistently be set to zero,
i.~e.~in terms of the above parametrization (29)
\begin{equation}
F_7(r,\theta)=0 \ , \ \ \ H_8(r,\theta)=0 \ , \ \ \ H_9(r,\theta)=0
\ . \end{equation}

The general spherically symmetric ansatz,
necessary to obtain the sphaleron barrier and the bisphalerons,
is given by
\begin{eqnarray}
  W_i^a & = & \frac{1-f_A(r)}{gr} \epsilon_{aij}\hat r_j
  + \frac{f_B(r)}{gr} (\delta_{ia}-\hat r_i \hat r_a)
  + \frac{f_C(r)}{gr} \hat r_i \hat r_a  \ , \nonumber \\
  W_0^a & = & 0 \ , \\
    \Phi & = & \frac{v}{\sqrt {2}}
  \Bigl(H(r) + i \vec \tau \cdot \hat r K(r)\Bigr)
    \left( \begin{array}{c} 0\\1  \end{array} \right)
\ . \end{eqnarray}
Comparing with the general axially symmetric ansatz
we find $n = 1$ and
\begin{eqnarray}
      F_1(r,\theta) &=& F_2(r,\theta) = F_3(r,\theta)
       = F_4(r,\theta) = { {1 - f_A(r)} \over 2} \ ,
\nonumber \\
      F_5(r,\theta) &=& F_6(r,\theta) = K(r) \ ,
\nonumber \\
      H_1(r,\theta) &=& H_2(r,\theta) = H_3(r,\theta)
       = { {f_C(r) - f_B(r)} \over 2} \ ,
\nonumber \\
      H_4(r,\theta) &=& H_5(r,\theta) = { f_B(r) \over 2} \ ,
\nonumber \\
      H_6(r,\theta) &=& 0 \ , \ \ \ H_7(r,\theta) = H(r)
\ . \end{eqnarray}
The functions $f_B(r), f_C(r)$ and (in the usual parametrization)
$H(r)$ represent the parity violating terms,
present in the configurations along the sphaleron barrier and
the bi\-sphalerons,
which generalize to seven functions $H_i(r,\theta)$
in the axially symmetric ansatz.
The spherically symmetric ansatz has a residual gauge symmetry,
which can be fixed for instance by requiring $f_C(r)=0$.

The spherically symmetric parity conserving sphaleron solution
has $f_B(r)=f_C(r)=H(r)=0$,
corresponding to the vanishing of all functions $H_i(r,\theta)$.

\subsection{Solutions with mirror symmetry}

Besides being axially symmmetric,
the sphaleron at finite mixing angle \cite{KKB1,KKB2},
the multisphalerons \cite{KK}
and the sphaleron-antisphaleron pair \cite{Kl1,Kl2}
have discrete symmeties.

Supplementing the axial invariance of the fields
by the discrete mirror symmetry
\begin{equation}
  M_{xz} \otimes   C \otimes (-\1op)_{\rm custodial}
\ , \end{equation}
where the first factor represents reflection through the $xz$-plane,
and the second factor denotes charge conjugation
\begin{equation}
    W_{\mu}^c = - W_{\mu}^T \ , \quad
    \Phi^c = \Phi^* \ , \quad
    A_{\mu}^c = - A_{\mu}
\ , \end{equation}
leads to the following simplifying conditions
\cite{Man1,RR,KKB1,KKB2}
\begin{equation}
 w_1^1 = w_1^2 = w_2^1 = w_2^2 = w_3^3 = 0
\ , \end{equation}
\begin{equation}
 h_3 = h_0 = 0
\ , \end{equation}
\begin{equation}
 a_1 = a_2 = 0
\ , \end{equation}
corresponding to
$H_i(r,\theta) = 0 \ , \ i=1,...,9$.

The known axially symmetric solutions are additionally
invariant under parity.

\subsubsection{Sphaleron and multisphalerons}

The sphaleron at finite mixing angle
and the multisphalerons
are described by the seven axial functions
$F_i(r,\theta)$ of Eqs.~(29) \cite{KKB1,KKB2,KK}.
The sphaleron and multisphaleron functions satisfy
\begin{equation}
  F_a (r , \theta) = F_a(r , \pi - \theta)
\ . \end{equation}
The solutions are invariant
under $P\otimes - \1op_{\rm custodial}$,
where the second factor is necessary because the classical Higgs field
is parity odd in the gauge used.

The Higgs fields of the sphaleron and of the multisphalerons ($S$)
assume the following asymptotic forms
\begin{equation}
 \Phi_S =  \frac{v}{\sqrt{2}} U_S(\infty)
    \left( \begin{array}{c} 0\\1  \end{array} \right)
        =i \frac{v}{\sqrt{2}}
           \Bigl(\sin \theta G_1^{(n)} (\phi) +
                 \cos \theta G_2^{(n)} (\phi) \Bigr)
    \left( \begin{array}{c} 0\\1  \end{array} \right)
\ . \end{equation}
The gauge fields become pure gauge configurations
at infinity
\begin{equation}
W_i(\infty) = -\frac{2i}{g} \partial_i U_S(\infty) U_S^\dagger(\infty)
\ . \end{equation}

Thus the boundary conditions for the functions $F_i(r,\theta)$ are
\cite{KKB1,KKB2,KK}
\begin{eqnarray}
r=0 & :          & \ \ F_i(r,\theta)|_{r=0}=0,
               \ \ \ \ \ \ i=1,...,7 \ ,
\nonumber \\
r\rightarrow\infty& :
	       & \ \ F_i(r,\theta)|_{r=\infty}=1,
               \ \ \ \ \ i=1,...,6,
               \ \ \ F_7(r,\theta)|_{r=\infty}=0 \ ,
\nonumber \\
\theta=0& :      & \ \ \partial_\theta F_i(r,\theta)|_{\theta=0} =0,
	       \ \ \ i=1,...,7 \ ,
\nonumber \\
\theta=\pi/2& :  & \ \ \partial_\theta F_i(r,\theta)|_{\theta=\pi/2} =0,
               \ i=1,...,7
\ . \end{eqnarray}

\subsubsection{Sphaleron-antisphaleron pair}

The sphaleron-antisphaleron pair
\cite{Kl1,Kl2} is also axially symmetric
and parity invariant.
But in contrast to the sphaleron the Higgs field is even under parity.

Klinkhamer denoted the field components as follows
\begin{eqnarray}
               w_1^3 &=& -{\alpha_1 \over \rho} \ , \quad \,
               w_2^3  =  -{\alpha_0 \over z   } \ , \quad
               w_3^1  =   {\alpha_2 \over \rho} \ , \quad
               w_3^2  =   {\alpha_3 \over \rho} \ ,
\nonumber \\
                h_1  &=&  \beta_1  \ , \quad \ \ \ \,
                h_2  =  - \beta_2  \ ,
\nonumber \\
                a_3  &=& {g'^2 \over g^2} {\alpha_4 \over \rho}
\ . \end{eqnarray}
He parametrized the gauge field components
in terms of the angle $\theta$ analogous to Eqs.~(29),
leading to the relations for the gauge field functions
\begin{equation}
F_1=-2\frac{r^2}{a}f_1^{\rm Kl} \ , \quad
F_2= 2\frac{r^2}{a}f_0^{\rm Kl} \ , \quad
F_3=-2\frac{r^2}{a}f_2^{\rm Kl} \ , \quad
F_4= 2\frac{r^2}{a}f_3^{\rm Kl}
\   \end{equation}
with $a=r^2+r_a^2$, and $r_a$ an arbitrary scale parameter,
while he parametrized the Higgs field components differently
\begin{equation}
h_1= \frac{r^2}{a} h_1^{\rm Kl} \sin 2\theta \ , \quad
h_2= h_2^{\rm Kl}
\ . \end{equation}
This parametrization lead to $\theta$-dependent boundary
conditions for the functions $f_2^{\rm Kl},
f_3^{\rm Kl}$ and $h_2^{\rm Kl}$.

The Higgs field of the sphaleron-antisphaleron pair ($S^*$)
assumes the asymptotic form
\begin{equation}
 \Phi_{S^*} =  \frac{v}{\sqrt{2}} U_{S^*}(\infty)
    \left( \begin{array}{c} 0\\1  \end{array} \right)
            =i \frac{v}{\sqrt{2}}
           \Bigr(\sin 2\theta G_1^{(1)} (\phi) +
                 \cos 2\theta G_2^{(1)} (\phi) \Bigr)
    \left( \begin{array}{c} 0\\1  \end{array} \right)
\ , \end{equation}
while the gauge fields become pure gauge configurations
\begin{equation}
W_i(\infty) = -\frac{2i}{g} \partial_i U_{S^*}(\infty)
                                       U_{S^*}^\dagger(\infty)
\ . \end{equation}
Therefore another parametrization appears to be natural
\begin{eqnarray}
       w_1^3 &=& { 4 \over gr} \tilde F_1(r,\theta) \cos \theta \ , \ \ \ \ \
\ \ \ \ \ \ \ \ \ \ \ \ \
       w_2^3  =-{4 \over gr} \tilde F_2(r,\theta) \sin \theta \ ,
\nonumber \\
       w_3^1 &=&-{4 \over gr} \tilde F_3(r,\theta) \cos \theta \cos
2\theta \ , \ \ \ \ \ \ \ \,
       w_3^2  = { 4\over gr} \tilde F_4(r,\theta) \cos \theta \sin 2\theta \ ,
\nonumber \\
       h_1 &=& \tilde F_5(r, \theta) \sin 2\theta \ , \ \ \ \ \ \ \ \ \ \ \ \
\ \ \ \ \ \ \ \ \
       h_2 = \tilde F_6(r, \theta) \cos 2\theta
\ . \end{eqnarray}
In terms of this parametrization the functions
$\tilde F_i(r,\theta), \ i=1,...,6$,
approach one at infinity.

\subsubsection{Generalization of the sphaleron-antisphaleron pair
ansatz}

Generalizing the ansatz for the sphaleron-antisphaleron pair
to arbitrary integers $m$,
we require for the Higgs field the asymptotic form
\begin{equation}
 \Phi_{S^*_m} =  \frac{v}{\sqrt{2}} U_{S^*_m}(\infty)
    \left( \begin{array}{c} 0\\1  \end{array} \right)
            =i \frac{v}{\sqrt{2}}
           \Bigr(\sin m\theta G_1^{(1)} (\phi) +
                 \cos m\theta G_2^{(1)} (\phi) \Bigr)
    \left( \begin{array}{c} 0\\1  \end{array} \right)
\ , \end{equation}
and for the gauge fields the pure gauge configurations
\begin{equation}
W_i(\infty) = -\frac{2i}{g} \partial_i U_{S^*_m}(\infty)
                                       U_{S^*_m}^\dagger(\infty)
\ , \end{equation}
leading to the general parametrization
\begin{eqnarray}
       w_1^3 &=& { 2m \over gr} \tilde F_1(r,\theta) \cos \theta \ , \ \ \ \ \
\ \ \ \ \ \ \ \ \ \ \ \ \ \
       w_2^3  =-{ 2m\over gr} \tilde F_2(r,\theta) \sin \theta \ ,
\nonumber \\
       w_3^1 &=&-{2m \over gr} \tilde F_3(r,\theta)
       \frac{\sin m \theta}{m \sin\theta}
       \cos m\theta \ , \ \ \ \ \
       w_3^2  = { 2m\over gr} \tilde F_4(r,\theta)
       \frac{\sin m \theta}{m \sin\theta}
       \sin m\theta \ ,
\nonumber \\
       h_1 &=& \tilde F_5(r, \theta) \sin m\theta \ , \ \ \ \ \ \ \ \ \ \ \ \
\ \ \ \ \ \ \ \ \ \
       h_2 = \tilde F_6(r, \theta) \cos m\theta
\ . \end{eqnarray}
In terms of this parametrization the functions
$\tilde F_i(r,\theta), \ i=1,...,6$,
approach one at infinity.

One further step is to include both integers $n$ and $m$
in the ansatz, i.~e.~use the gauge transformation $U_{S_{n,m}}$
for the fields at infinity
\begin{equation}
 \Phi_{S_{n,m}} =  \frac{v}{\sqrt{2}} U_{S_{n,m}}(\infty)
    \left( \begin{array}{c} 0\\1  \end{array} \right)
            =i \frac{v}{\sqrt{2}}
           \Bigr(\sin m\theta G_1^{(n)} (\phi) +
                 \cos m\theta G_2^{(n)} (\phi) \Bigr)
    \left( \begin{array}{c} 0\\1  \end{array} \right)
\ , \end{equation}
and
\begin{equation}
W_i(\infty) = -\frac{2i}{g} \partial_i U_{S_{n,m}}(\infty)
                                       U_{S_{n,m}}^\dagger(\infty)
\ . \end{equation}

\section{Chern-Simons charge}

The Chern-Simons current $K_\mu$
is not conserved,
its divergence $\partial^\mu K_\mu$
represents the U(1) anomaly
of the baryon current.
Classical configurations are
characterized by their Chern-Simons charge.
The SU(2) part of the Chern-Simons charge is given by
\begin{eqnarray}
N_{\rm CS} = \int d^3r K^0
  &=& - {g^2 \over 64 \pi^2} \int d^3 r
     \epsilon_{ijk} {\rm Tr}
\Bigl(F_{ij} W_k +i {g\over 3} W_i W_j W_k \Bigr)
\nonumber \\
          &=& {1 \over 2 \pi^2} \int d^3 r Q(\rho,z)
\ . \end{eqnarray}

The proper gauge for evaluating the Chern-Simons charge is
the gauge, where the gauge field is given by
\begin{equation}
W_i(\infty) = -\frac{2i}{g} \partial_i U(\infty) U^\dagger(\infty)
\ , \end{equation}
with $U(\infty)=1$.
Then this Chern-Simons charge of the configurations corresponds to their
baryonic charge, when the U(1) field does not
contribute to the baryon number [5].

\subsection{General axially symmetric ansatz}

The general axially symmetric ansatz leads to a Chern-Simons charge
characterized by
\begin{eqnarray}
  -\frac{4}{g^3} Q(\rho,z) &=&  w_1^1 w_2^2 w_3^3 + w_1^2 w_2^3 w_3^1
  + w_1^3 w_2^1 (w_3^2-\frac{n}{g\rho}) \nonumber \\
       &- & w_1^1 w_2^3 (w_3^2-\frac{n}{g\rho}) -
  w_1^3 w_2^2 w_3^1 - w_1^2 w_2^1 w_3^3  \\
 &-& {1 \over g} \Bigl(w_3^1 (\partial_z w_1^1 - \partial_{\rho} w_2^1)
                 +(w_3^2-\frac{n}{g\rho})
                  (\partial_z w_1^2 - \partial_{\rho} w_2^2)
                 +w_3^3 (\partial_z w_1^3 - \partial_{\rho} w_2^3)
\Bigr) \ , \nonumber
\ . \end{eqnarray}
This expression must be supplemented
by the appropriate gauge transformation
to obtain the Chern-Simons charge
of the configurations
forming the sphaleron barrier at finite mixing angle
and the multisphaleron barriers,
and to obtain the Chern-Simons charge of the bisphalerons.

\subsection{Sphaleron and multisphalerons}

For the sphaleron and multisphalerons
the Chern-Simons density is proportional to
\begin{eqnarray}
Q(\rho,z) &=&
      n { \sin^2 \Omega \over r^2}
        {\partial \Omega \over \partial r}
\nonumber \\
 &+& n {\partial  \over \partial z}
    \Bigl( { z \over 4 r^3} F_1 \sin (2 \Omega) \Bigr)
  +{n\over \rho} {\partial  \over \partial \rho}
    \Bigl( { \rho ^2 \over 4 r^3} F_2 \sin (2 \Omega) \Bigr)
\nonumber \\
 &+& {\cos^2 \theta \over 4r^2} {\partial \over \partial r}
    \Bigl( F_3 \sin (2 \Omega)  \Bigr)
  +  {\sin^2 \theta \over 4r^2} {\partial \over \partial r}
    \Bigl( F_4 \sin (2 \Omega)  \Bigr)
\nonumber \\
 &+& {1\over 2 \rho} {\partial \over \partial z}
   \Bigl( {{z \rho^2}\over r^3} (F_3 - F_4)
   {\partial \Omega \over \partial \rho} \Bigr)
  -  {1\over 2 \rho} {\partial \over \partial \rho}
   \Bigl( {{z \rho^2}\over r^3} (F_3 - F_4)
   {\partial \Omega \over \partial z} \Bigr)
\ , \end{eqnarray}
where, analogous to Ref.~\cite{KM},
we incorporated the effect of a gauge transformation of the form
\begin{equation}
U_S(\vec r\,) = \exp
    \Bigl( i\Omega(r,\theta)
    (\sin \theta G_1^{(n)} +\cos \theta G_2^{(n)}) \Bigr)
\   \end{equation}
and kept all derivative terms.
The proper boundary conditions are $\Omega(0) = 0$ and
$\Omega(\infty) = \pi /2$ \cite{KM}.
Only the first term of $Q(\rho,z)$ determines the Chern-Simons charge,
since the derivative terms do not contribute
due to the boundary conditions for the functions $F_i(r,\theta)$
(see Eqs.~(42) \cite{KKB1,KKB2,KK}) and for $\Omega(r,\theta)$.

We find for the multisphalerons the Chern-Simons charge
\begin{equation}
N_{\rm CS} = n/2
\ , \end{equation}
independently of the Higgs mass and of the mixing angle,
reproducing the well-known Chern-Simons charge of the sphaleron,
$N_{\rm CS} = 1/2$.
This Chern-Simons charge of the sphaleron and of
the multisphalerons corresponds to their
baryonic charge, $Q_B = n/2$, since the U(1) field does not
contribute to their baryon number [5].

\subsection{Generalization of the sphaleron-antisphaleron pair ansatz}

For the sphaleron-antisphaleron pair another gauge transformation
must be chosen to evaluate the Chern-Simons charge
\begin{equation}
U_{S^*}(\vec r\,) = \exp
    \Bigl( i\Omega(r,\theta)
    (\sin 2\theta G_1^{(1)} +\cos 2\theta G_2^{(1)}) \Bigr)
\ , \end{equation}
with boundary conditions
$\Omega(0)=0$ and $\Omega(\infty)=\pi/2$,
since this solution approaches infinity differently.

In the following we present the
Chern-Simons density directly for the generalized pair ansatz
Eqs.~(49)-(51), using the gauge transformation
\begin{equation}
U_{S^*_m}(\vec r\,) = \exp
    \Bigl( i\Omega(r,\theta)
    (\sin m\theta G_1^{(1)} +\cos m\theta G_2^{(1)}) \Bigr)
\ . \end{equation}
The Chern-Simons density is proportional to
\begin{eqnarray}
Q(r,\theta) &=&
       \frac{m\sin m\theta}{\sin \theta}
        { \sin^2 \Omega \over r^2}
        {\partial \Omega \over \partial r}
\\
 &+&  {\partial  \over \partial z}
    \Bigl(
    \frac{m\sin m\theta}{\sin \theta}
    \frac{ \cos \theta }{4 r^2}
     \tilde F_1 \sin (2 \Omega) \Bigr)
  +{1\over \rho} {\partial  \over \partial \rho}
    \Bigl(
    \frac{m\sin m\theta}{\sin \theta}
    { \sin^2 \theta \over 4 r}
    \tilde F_2 \sin (2 \Omega) \Bigr)
\nonumber \\
 &+& \frac{m\sin m\theta}{\sin \theta} \Biggl(
     {\cos^2 m\theta \over 4r^2} {\partial \over \partial r}
    \Bigl( \tilde F_3 \sin (2 \Omega)  \Bigr)
  +  {\sin^2 m\theta \over 4r^2} {\partial \over \partial r}
    \Bigl( \tilde F_4 \sin (2 \Omega)  \Bigr) \Biggr)
\nonumber \\
 &+& {1\over 2 \rho} {\partial \over \partial z}
   \Bigl( \cos m \theta \sin^2 m \theta
   (\tilde F_3 - \tilde F_4) {\partial \Omega \over \partial \rho} \Bigr)
  -  {1\over 2 \rho} {\partial \over \partial \rho}
   \Bigl( \cos m \theta \sin^2 m \theta
   (\tilde F_3 - \tilde F_4) {\partial \Omega \over \partial z} \Bigr) \ ,
\nonumber
\   \end{eqnarray}
where we kept all derivative terms.
With the proper boundary conditions
for the functions $\tilde F_i$, and
for the gauge function, $\Omega(0) = 0$ and $\Omega(\infty) = \pi /2$,
again only the first term of $Q(\rho,z)$
determines the Chern-Simons charge,
since the derivative terms do not contribute.
We find the Chern-Simons charge
\begin{equation}
   N_{\rm CS} = \frac{1-\cos m \pi}{4} = \left\{ \begin{array}{ll}
  \frac{1}{2} & \mbox{if $m$ odd} \\
  0           & \mbox{if $m$ even} \end{array} \right.
 \ . \end{equation}

\section{Conclusions}

We have presented the general ansatz,
the energy density and the Chern-Simons charge
for static axially symmetric configurations
in the bosonic sector of the electroweak theory.

The ansatz contains the known axially symmetric solutions
with parity reflection symmetry,
the sphaleron, the multisphalerons
and the sphaleron-antisphaleron pair at finite mixing angle.
It further allows for the construction
of configurations without parity reflection symmetry,
such as the sphaleron and multisphaleron barriers at finite mixing angle
and the bisphalerons at finite mixing angle.
The leading correction to the spherical bisphalerons
was obtained in a perturbative calculations in $\theta_{\rm w}$
\cite{BK}.
The change of the sphaleron barrier due to the finite mixing angle
as well as the barriers associated with the multisphalerons
have not yet been obtained.
The construction of the multisphaleron barriers will allow
the investigation of the fermion level crossing phenomenon
for vacuum to vacuum transitions via multisphalerons.

The numerical construction of these barriers or
of the bisphalerons at finite mixing angle
now appears to be straightforward, at least in the Coulomb gauges,
but numerically involved, because a large system of
up to 16 partial non-linear differential equations must be solved
simultaneously.

The multisphalerons are characterized by an integer winding number $n$,
describing the winding the fields with
respect to the angle $\phi$.
Their Chern-Simons charge is given by $N_{\rm CS}=n/2$.
The sphaleron has winding number $n=1$.
Since the bisphalerons bifurcate from the sphaleron at large
Higgs masses, we expect that corresponding $n$-bisphalerons exist,
bifurcating from the multisphalerons with winding number $n$.
Using the formalism derived in this paper,
these solutions can numerically be searched for.
A stability analysis of the multisphaleron solutions
may be helpful in determining the critical values of the Higgs mass.

Besides the winding in the angle $\phi$ a winding in the angle
$\theta$ with winding number $m$ can be considered.
The sphaleron-antisphaleron pair represents a solution
with winding number $m=2$.
We have generalized the ansatz for the sphaleron-antisphaleron pair
to allow for arbitrary integer winding number $m$.
The Chern-Simons charge of solutions with odd $m$ is
$N_{\rm CS}=1/2$,
while the Chern-Simons charge of solutions
with even $m$ vanishes, $N_{\rm CS}=0$.
The sphaleron has winding number $m=1$.
We conjecture that solutions with winding number $m>2$ exist.
Further there may be solutions with both winding numbers excited,
$n>1$ and $m>1$. The numerical construction of such solutions
may turn out to be complicated, though only seven functions are
involved.

Finally, all these solutions may bifurcate and general bisphalerons
with $m>1$ and with
$n>1$ and $m>1$ may exist for large Higgs masses.
The construction of such solutions provides
a great numerical challenge.

\vfill \eject

\vfill\eject
\end{document}